\documentclass[journal]{IEEEtran}

\usepackage{cite}
\usepackage{amsmath,amssymb,amsfonts}
\usepackage{graphicx}
\usepackage{esint}
\usepackage{booktabs}
\usepackage{etoolbox,siunitx}
\robustify\bfseries
\usepackage{cite}
    
\begin{document}
\sisetup{detect-weight=true,detect-inline-weight=math}

\title{Magnet system for the Quantum Electro-Mechanical Metrology Suite}

\author{Rafael~R.~Marangoni, Darine~Haddad, Frank~Seifert, Leon~S.~Chao, David~B.~Newell, Stephan~Schlamminger
\thanks{R.~R.~Marangoni, D.~Haddad, F.~Seifert, L.~S.~Chao, D.~B.~Newell and S.~Schlamminger are with the National Institute of Standards and Technology, Gaithersburg, MD 20899, United States of America.}
\thanks{F.~Seifert is with the University of Maryland, Joint Quantum Institute, College Park, MD 20742, United States of America.}}

\maketitle

\begin{abstract}
The design of the permanent magnet system for the new Quantum Electro-Mechanical Metrology Suite (QEMMS) is described. The QEMMS, developed at the National Institute of Standards and Technology (NIST), consists of a Kibble balance, a programmable Josephson voltage standard, and a quantum Hall resistance standard. It  will be used to measure masses up to \SI{100}{g} with relative uncertainties below \SI{2e-8}{}. The magnet system is based on the design of the NIST-4 magnet system with significant changes to adopt to a smaller Kibble balance and to overcome known practical limitations.
Analytical models are provided to describe the coil-current effect and model the forces required to split the magnet in two parts to install the coil. Both models are compared to simulation results obtained with finite element analysis and measurement results. Other aspects, such as the coil design and flatness of $\boldsymbol{Bl}$ profile are considered.
\end{abstract}


\section{Introduction}
\IEEEPARstart{T}{he} Fundamental Electrical Measurements Group of the National Institute of Standards and Technology (NIST) is developing the Quantum Electro-Mechanical Metrology Suite (QEMMS). This suite is a quantum metrology solution composed of a Kibble balance, a programmable Josephson voltage standard and quantum Hall resistance standard. The Kibble balance is being designed for measuring masses of \SI{100}{g} with relative uncertainties lower than \SI{2e-8}{}.
Since the revision of the International System of Units (SI) on 20th May 2019, the definition of the kilogram unit is based on the Planck constant $h$ and the definitions of the meter and second via fixed values of the speed of light and the unperturbed ground-state hyperfine transition frequency of Caesium-133, respectively. The Kibble balance provides one way to realize the unit of mass at the kilogram level with high accuracy.

In a Kibble balance, the weight of a mass $m$ in a region with acceleration of free fall $g$ is compensated by the electromagnetic force of an electric current $I$ in a magnetic field with flux density $B$. The electric current flows in a multi-turn coil with length $l$. The high accuracy of the mass measurement can be achieved by precisely measuring $g$, $I$ and the product $Bl$ during balance operation. The acceleration of free fall is measured with an absolute gravimeter, the electric current is measured by using a reference resistor in combination with a voltmeter and the product $Bl$ is measured by moving the coil in the magnetic field and determining the ratio between induced voltage and velocity.
With the Kibble balance it is theoretically possible to calibrate standards of any mass value directly. This is an advantage for masses lower than \SI{1}{\kilo\gram}, which can be measured with lower uncertainties by reducing the metrological traceability chain \cite{Stock_2019}.
 
The magnet system of a Kibble balance is responsible for generating the magnetic field with flux density ${B}$.
Usually a magnetic circuit with permanent magnets in an axially symmetric configuration is used to generate the magnetic field.
The field must be stable in time and have a uniform profile along the travel range of the coil.
It is advantageous to have the magnetic flux density with the highest possible magnitude.
The requirements and design aspects for the magnet system of the QEMMS are described in this article.

\begin{figure}
    \centering
    \includegraphics{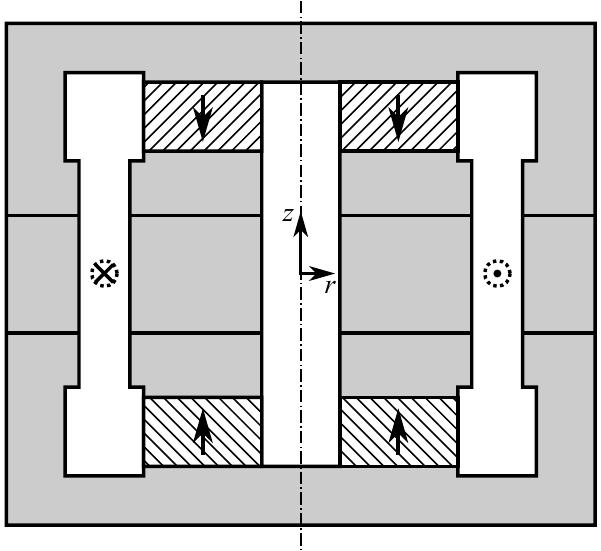}
    \caption{Drawing of the magnet system. The hatched areas are the permanent magnets with the magnetization directions indicated by the arrows. The gray areas represent the soft iron yoke. This geometry was used originally by the BIPM and has been copied by various laboratories around the world, including NIST. The assembly is similar to the NIST-4 magnet \cite{Schlamminger_2012} and the yoke is composed by 6 parts of soft steel.}
    \label{fig:drawing_maget_design}
\end{figure}

\section{Description of the system}

The idea of the QEMMS Kibble balance is to realize the kilogram unit for masses up to \SI{100}{\gram} with relative uncertainties smaller than \SI{2e-8}{} and as few operational requirements as possible. When compared to the NIST-4, the QEMMS will be smaller, simpler to operate, and easier to maintain.
These principles will be used in the design of the balance components, including the magnet system.

Several requirements for the magnet system were defined prior to the design based on experiences gained with NIST-4 and Kibble balances at other laboratories: (1) the precision air gap should have a height of \SI{4}{\centi\metre}, (2) the maximum relative deviation of the radial flux density in the precision air gap should be below \SI{1e-4}{}, and (3) the temperature coefficient of the flux density should be \SI{1e-5}{\kelvin^{-1}} or smaller. While the allowed variation of the flux density of the field in the precision gap is similar to that of NIST-4, the precision air gap height is two times and the temperature sensitivity 33 times smaller than the corresponding parameter of the magnet system used for the NIST-4~\cite{Seifert_2014}.

As described in \cite{Schlamminger_2012}, there are basically five possibilities for generating the magnetic flux density $B$: conventional electromagnets, superconducting magnets, permanent magnets with yoke, yokeless permanent magnets and hybrid magnets that combine any of these options. There are advantages and disadvantages of these possibilities for generating $B$.
The permanent magnets represent an option with relative simple design and operation, while having low cost and  maintenance. There are different designs for magnet systems based on permanent magnets used in Kibble balances \cite{Robinson_2016}.
The design pioneered by the International Bureau of Weights and Measures (BIPM) is the most frequently used and is employed by several laboratories around the world, including the BIPM itself and the national metrology institutes in China \cite{Li_2019}, Germany \cite{Rothleitner_2018}, South Korea \cite{Kim_2017}, Switzerland \cite{Tommasini_2016}, Turkey \cite{Ahmedov_2018} and United States \cite{Schlamminger_2012}.
Figure~\ref{fig:drawing_maget_design} shows a drawing of the magnet system, which is based on the NIST-4 Kibble balance.
Due to several advantages of the BIPM design, this option was chosen for the QEMMS Kibble balance.
This design provides good shielding due to the closed yoke and has a great past performance in previous versions of Kibble balances. 
This magnet system was used during high-precision measurements of the Planck constant that occurred before the redefinition of the SI \cite{Haddad_2017}.
The main drawback of the design is that the magnet must be  split apart to install, access, or repair the coil.
For the NIST-4 magnet, the force required to split the magnet is about \SI{4.7}{\kilo\newton}. Such a large force makes the splitting process difficult and a dedicated magnet splitter was necessary~\cite{Seifert_2014}.
The large mass of the NIST-4 magnet system of about \SI{850}{\kilo\gram} makes the manipulation of the magnet system even more complicated. 

\begin{figure}
    \centering
    \includegraphics{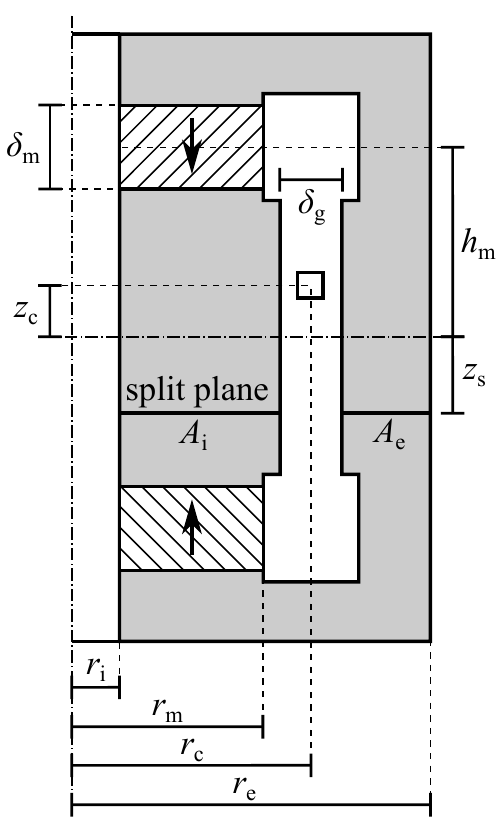}
    \caption{Dimensions of the magnet system used for the analytical models. There are 9 main geometrical parameters present in the models described here.}
    \label{fig:dimensions_magnet_system}
\end{figure}

Both disadvantages, a heavy magnet and the need for a dedicated splitter, were addressed in the design for the QEMMS magnet. As a result, the discussed magnet weighs only \SI{110}{kg} and can be split by applying a reasonable force of \SI{250}{\newton}. The latter allows the integration of the magnet splitter into the magnet design, eliminating the need for a dedicated device. Similar to the magnet in the NIST-4, mounting fixtures will be bolted on the upper part of the magnet. As shown in figure~\ref{fig:dimensions_magnet_system}, the lower third of the magnet can be removed. To further simplify the split process, the magnet system was designed such that the weight of the lower third matches the split force. Hence, it is not necessary to generate any additional force during the split process.

Several aspects must be considered designing a magnet system \cite{Schlamminger_2012, Li_2013, Seifert_2014, Li_2017, Li_2017_2}: (1) the back-action of the current on the magnetic flux density, (2) the flatness of the flux density profile,  (3) the temperature dependency, and (4) the geometrical requirements of the balance, i.e., coil radius and length of the flat region. 
To design a magnet system that fulfills all requirements it is necessary to evaluate several aspects using analytic models and simulations. The analysis is described in the next section. 

\section{Performance evaluation}

Five aspects of the magnet system are evaluated. These are: (1) the magnetic circuit, (2) the reluctance force on the coil, (3) the force required to split the magnet,  (4) the flatness of the flux density in the air gap  as a function of vertical position, i.e., the field profile, and (5) the required size, wire gauge, and the number of turns for the coil. For this evaluation, the dimensions shown in figure~\ref{fig:dimensions_magnet_system} were used and are summarized as follows: width of the air gap $\delta_\mathrm{g}=\SI{2.6}{\centi\metre}$, height of each permanent magnet $\delta_\mathrm{m}=\SI{3.5}{\centi\metre}$, distance of the center of the permanent magnet from the symmetry plane $h_\mathrm{m}=\SI{8}{\centi\metre}$, location of the split plane relative to the symmetry plane $z_\mathrm{s}= \SI{3}{\centi\metre}$, radius of the internal bore $r_\mathrm{i}=\SI{2}{\centi\metre}$, radius of the permanent magnet  $r_\mathrm{m}=\SI{8}{\centi\metre}$, radius of the coil $r_\mathrm{c}=\SI{10}{\centi\metre}$ and outer radius of the complete system $r_\mathrm{e}=\SI{15}{\centi\metre}$. The active magnetic components of the magnet system are two composite, identical rings made from TC-16 ($\mathrm{Sm_{2}Co_{17}Gd}$) arc segments. Alloying Gd together with SmCo reduces the remanence temperature coefficient of the magnet  to \SI{-0.001}{\percent\per\kelvin}. The price to pay for the temperature compensation is a reduced remanence  of $\mathrm{Sm_{2}Co_{17}Gd}$ compared to that of  $\mathrm{Sm_{2}Co_{17}}$. For the material used the remanence is $B_\mathrm{r}$ of \SI{0.83}{\tesla}.

\subsection{Magnetic circuit}

The magnetic circuit equation is used to estimate the magnetic flux density in the air gap. In Appendix~\ref{appx:magnetic_circuit}, a derivation of the magnetic flux $\Phi$ through the coil based on the magnetic circuit is provided. The product $Bl$ and the magnetic flux are related by the following expression:
\begin{equation} \label{eq:Bl_mag_flux}
    Bl=N\frac{\mathrm{d}\Phi}{\mathrm{d}z_\mathrm{c}}(I=0)
\end{equation}

By combining equations (\ref{eq:Bl_mag_flux}) and (\ref{eq:mag_flux}), and considering the fact that $l=2\pi r_\mathrm{c}N$, the following expression for the magnetic flux density $B$ in the air gap can be obtained:
\begin{equation}
    B = \frac{B_\mathrm{r}}{2r_\mathrm{c}h_\mathrm{m}/(r_\mathrm{m}^2-r_\mathrm{i}^2) + \mu_\mathrm{m}\delta_\mathrm{g}/\delta_\mathrm{m}}
    \label{eq:fluxcalc}
\end{equation}
where $\mu_\mathrm{m}=1.06$ is the relative permeability of the magnetic material. Using equation~\ref{eq:fluxcalc} the magnetic flux density in the air gap is estimated to be  \SI{240}{\milli\tesla}. Finite element analysis yields a value of approximately \SI{245}{\milli\tesla}. The same calculations were performed for the NIST-4 magnet system. The results from the analytical model and simulation are \SI{525}{\milli\tesla} and \SI{554}{\milli\tesla}, respectively. Measurements published in \cite{Seifert_2014} indicate a value of \SI{553}{\milli\tesla} for the radial flux density. Hence, there is a good agreement between model, simulation and measurement. The analytical model represents a simple way to determine the magnetic flux in the air gap as a function of the parameters of the magnet system.

\begin{figure}
    \centering
    \includegraphics{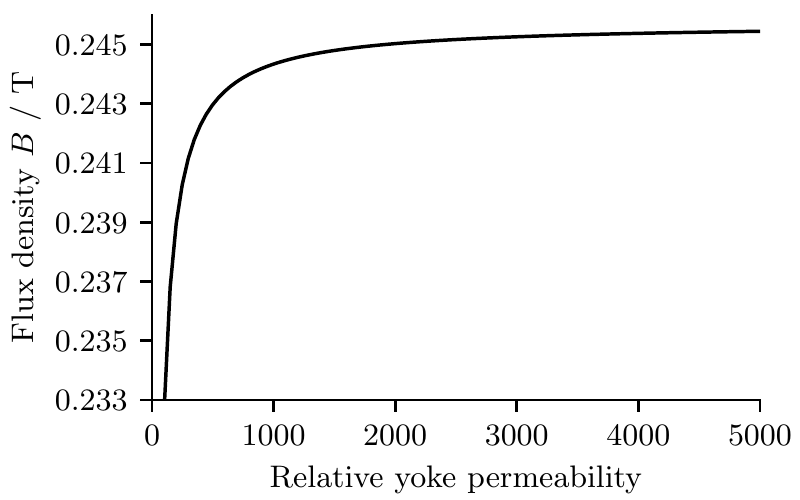}
    \caption{Radial flux density in the air gap as a function of the yoke permeability. The expected relative permeability of the material to be used is \SI{2700}{}.}
    \label{fig:permeability_change}
\end{figure}

For the determination of the magnet circuit equation, it was assumed that the yoke material has a high permeability. Soft steel classified by American Iron and Steel Institute (AISI) as 1010 is considered for the yoke, and a permeability of 2700 is expected. This material offers a high permeability for a reasonable cost. Figure~\ref{fig:permeability_change} shows the radial flux density in the air gap as a function of the yoke permeability. This figure is a simulation result obtained with finite element analysis. A significant variation in the flux density can be observed for a relative permeability smaller than 1000. For higher values of permeability, the variation of the flux density in the air gap is very small. The magnet system is designed such that the iron is not magnetically saturated at any point in the magnetic circuit, and variations in the permeability don't affect much the flux density in the air gap.

\subsection{Reluctance force} \label{sec:rel_force}

The reluctance force is caused by a position-dependent variation of the magnetic flux through the coil when an electric current is flowing. This problem has already been considered in \cite{Schlamminger_2012,Li_2017,Li_2017_2}. A similar analysis is performed here. In this analysis, the magnetic flux in both air gap and permanent magnet are considered to determine the reluctance force.

This force can be determined by using the following expression \cite{Kirtley_2010}:
\begin{equation} \label{eq:reluc_force_energy}
    F_\mathrm{r}=-\frac{\partial E_\mathrm{c}}{\partial z_\mathrm{c}}
\end{equation}
where $E_\mathrm{c}$ is the energy stored in the coil. It can be obtained by integrating the coil power $P_\mathrm{c}=\mathrm{d} E_\mathrm{c}/ \mathrm{d} t = u_\mathrm{c}I$, where the coil voltage is determined by using Faraday's law of induction:
\begin{equation}
    u_\mathrm{c}=-N\frac{\mathrm{d}\Phi}{\mathrm{d}t}=-N\frac{\partial\Phi}{\partial I}\frac{\mathrm{d}I}{\mathrm{d}t}
\end{equation}

The energy stored in the magnetic flux caused by the coil current is given by:
\begin{equation} \label{eq:energy_reluc_force}
    E_\mathrm{c} = \int \mathrm{d}E_\mathrm{c} = - N \int I\frac{\partial \Phi}{\partial I}\mathrm{d}I = -\frac{NI^2}{2}\frac{\partial \Phi}{\partial I}
\end{equation}

By combining the equations (\ref{eq:reluc_force_energy}), (\ref{eq:energy_reluc_force}) and (\ref{eq:mag_flux}), the following expression for the reluctance force can be obtained:
\begin{equation}
    F_\mathrm{r} = -\frac{\pi\mu_0 N^2  r_{\mathrm{c}} z_\mathrm{c} I^2  / \delta_{\mathrm{g}}}{ h_\mathrm{m} + \mu_\mathrm{m} \delta_{\mathrm{g}} (r_{\mathrm{m}}^2-r_{\mathrm{i}}^2) / (2r_{\mathrm{c}}\delta_{\mathrm{m}})} 
\end{equation}

The reluctance force can also be described relative to the nominal force generated by the coil, which is equal to $F=BlI$. The relative reluctance force is defined as $f_\mathrm{r}=F_\mathrm{r}/F$ and is given by:
\begin{equation}
    f_\mathrm{r} = -\frac{\mu_0  r_\mathrm{c} N I z_\mathrm{c} }{\delta_\mathrm{g} B_\mathrm{r} (r_\mathrm{m}^2 - r_\mathrm{i}^2)}
\end{equation}

For a given magnetic material, the relative reluctance force can be reduced by increasing the air gap width $\delta_\mathrm{g}$ or the radius of the magnetic material $r_\mathrm{m}$. By reducing the number of turns $N$, the coil radius $r_\mathrm{c}$, the internal hole radius $r_\mathrm{i}$ or the coil current $I$, the relative reluctance force can also be reduced. The ratio between the relative reluctance force and the coil position  multiplied by the current $(z_\mathrm{c}I)$ gives a constant value that can be used to quantify the influence of the reluctance force. The reluctance force constant is defined as:
\begin{equation}
    c_\mathrm{rf}=\frac{f_\mathrm{r}}{z_\mathrm{c}I}
\end{equation}

In order to avoid measurement deviations, it is important to minimize this quantity. The magnet system of the QEMMS Kibble balance gives a $c_\mathrm{rf}$ of approximately \SI{-4.51}{\per\metre\per\ampere}. A value of \SI{-4.45}{\per\metre\per\ampere} was obtained by performing a finite element analysis. For these calculations, a  $Bl$ of \SI{700}{\tesla\metre} was used. The reluctance force constant was also determined for the \mbox{NIST-4} magnet system, resulting in \SI{-0.237}{\per\metre\per\ampere} and \SI{-0.235}{\per\metre\per\ampere} for the model and finite element analysis respectively. Reference \cite{Seifert_2014} contains a measurement result of the second derivative of the coil inductance  with respect to vertical position. The value $\partial^2L/\partial z_\mathrm{c}^2 = \SI{-346}{\henry/\metre^2}$ can be converted to the reluctance force constant and yields \SI{-0.244}{\per\metre\per\ampere}. This conversion can be performed by using  $Bl= \SI{709}{\tesla\metre}$ and:
\begin{equation}
    c_\mathrm{rf}=\frac{1}{2Bl}\frac{\partial^2L}{\partial z_\mathrm{c}^2},
\end{equation}

The equation proposed here agrees better with simulation and measurement than previously published equations in \cite{Schlamminger_2012,Li_2017_2}. The improvement stems from the fact that here the entire magnetic flux through the coil is considered for the determination of the reluctance force, and not just the magnetic flux through the air gap.

In Kibble balance experiments, the weighing process is normally performed in a way to eliminate measurement deviations caused by the reluctance force. For this, the weighing process is divided in two phases, named mass-on and mass-off. For the first phase the measuring mass $m$ is placed on the weighing pan and, for the second phase, the same mass is removed. The following equations can be obtained for both phases:
\begin{equation} \label{eq:mass_on}
    BlI_\mathrm{On}(1+c_\mathrm{rf}z_\mathrm{c}I_\mathrm{On}) = mg - m_\mathrm{t}g
\end{equation}
\begin{equation} \label{eq:mass_off}
     BlI_\mathrm{Off}(1+c_\mathrm{rf}z_\mathrm{c}I_\mathrm{Off}) = -m_\mathrm{t}g
\end{equation}

The tare mass  $m_\mathrm{t}$ represents an imbalance added to the balance counterweight. It is adjusted such that the weighing currents have equal magnitude and opposite directions:
\begin{equation} \label{eq:current_on_off}
    I_\mathrm{On}=-I_\mathrm{Off}=\Delta I 
\end{equation}

By subtracting equation~(\ref{eq:mass_off}) from~(\ref{eq:mass_on}) and using~(\ref{eq:current_on_off}), the following expression is obtained:
\begin{equation} \label{eq:mass_det}
    m=\frac{2\Delta IBl}{g}
\end{equation}

In theory, the component related to the reluctance force drops out and is not present in equation~(\ref{eq:mass_det}). However, to obtain this result it was assumed that the coil position $z_\mathrm{c}$ is the same for both mass-on and mass-off phases. This is not the case in practice \cite{Haddad_2017} and, for this reason, it is important to minimize the reluctance force constant $c_\mathrm{rf}$ during the design of the magnet system.

\subsection{Force required to split the magnet}

Appendix~\ref{appx:split_force} contains the derivation of an analytical equation to calculate the force required to split the magnet as a function of the vertical position of the split plane. The final equation is obtained by combining the equations (\ref{eq:F_z}), (\ref{eq:B_ze}) and (\ref{eq:B_zi}), yielding
\begin{equation} \label{eq:split_force}
    F_\mathrm{z}=\frac{\pi r_{\mathrm{c}} B^2}{\mu_0} \left(  \frac{2\pi r_\mathrm{c} (A_\mathrm{i} + A_\mathrm{e})}{A_\mathrm{i} A_\mathrm{e}}z_\mathrm{s}^2 - \delta_{\mathrm{g}} \right),
\end{equation}
where $A_\mathrm{i}$ and $A_\mathrm{e}$ are the areas of the yoke in the split plane as indicated in figure~\ref{fig:dimensions_magnet_system}. They are given by the equations~(\ref{eq:area_ai}) and~(\ref{eq:area_ae}) respectively.

\begin{figure}
    \centering
    \includegraphics{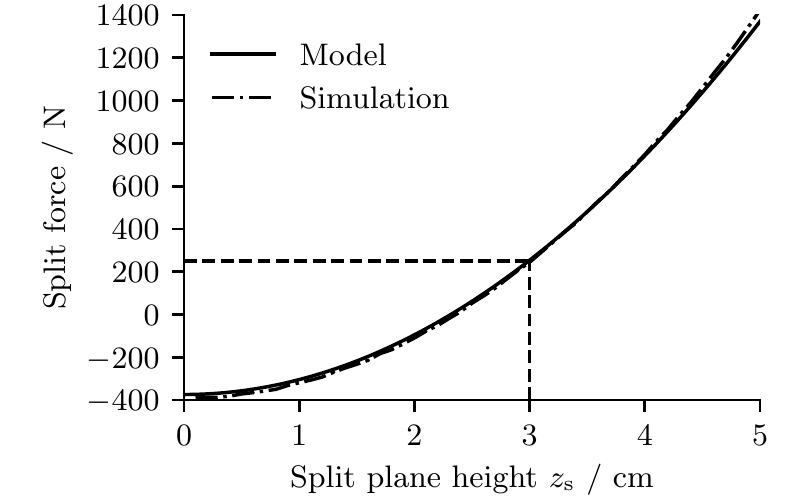}
    \caption{Model and simulation result for split force as a function of the split plane height. For the simulation, finite element analysis was used. Negative represents force in repulsive direction and positive represents force in attractive direction.}
    \label{fig:split_plane_force}
\end{figure}

\begin{figure}
    \centering
    \includegraphics{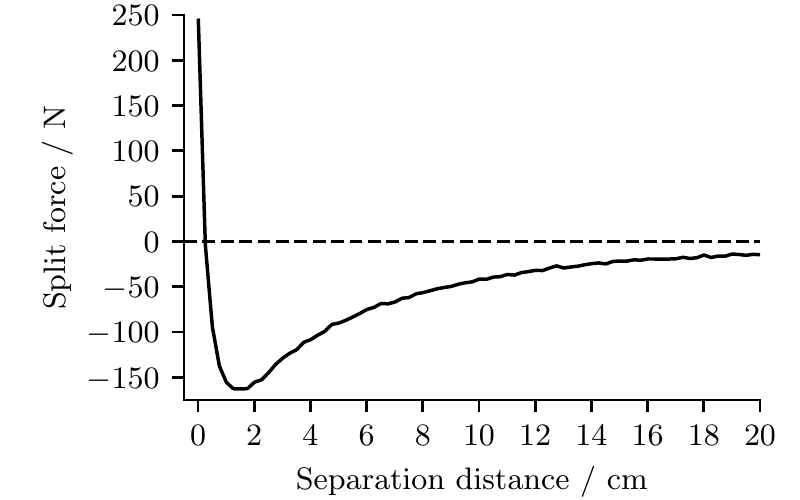}
    \caption{Simulation result for split force as a function of the separation between the magnets. This is a result of finite element analysis.  Negative represents force in repulsive direction and positive represents force in attractive direction.}
    \label{fig:split_force_separation}
\end{figure}

Figure~\ref{fig:split_plane_force} shows a comparison between equation~(\ref{eq:split_force}) and a simulation for the QEMMS Kibble balance magnet system. There is a good agreement between model and simulation results. For a split plane in the middle ($z_s=0$) , a repulsive force of about \SI{400}{\newton} is present. Choosing the location of the split plane at $z_s= \SI{2.4}{\centi\metre}$ results in zero split force. For higher split plane heights the force becomes repulsive and increases. By analyzing this plot, considering that the split plane should fall outside the precision air gap with a height of \SI{\pm2}{\centi\metre}, the location of the split plane was chosen to be at $z_s=\SI{3}{\centi\metre}$. For this height, the force required to open the magnet system is \SI{250}{\newton}. The mass of the lower third of the magnet system is estimated to be \SI{28}{\kilo\gram}. When suspended from the top, the weight of the lower third is equal and opposite the split force, simplifying the split operation. Since the additional force that is required to open the magnet is close to zero, the magnet splitter can be integrated into the magnet design instead of building a dedicated device.

Figure~\ref{fig:split_force_separation} shows a simulation result for the  force on either magnet part as a function of separation for a split plane height of $z_\mathrm{s} = \SI{3}{\centi\metre}$. The initial force  is attractive with a magnitude of \SI{250}{\newton}. At a distance of  \SI{0.25}{\centi\metre}, the force reverses sign and becomes repulsive. The largest repulsive force with a magnitude of about \SI{160}{\newton} is observed for a distance between \SI{1}{\centi\metre} and \SI{2}{\centi\metre}. For larger distances the force converges to \SI{0}{\newton}.

\subsection{Profile of the magnetic field}

\begin{figure}
    \centering
    \includegraphics{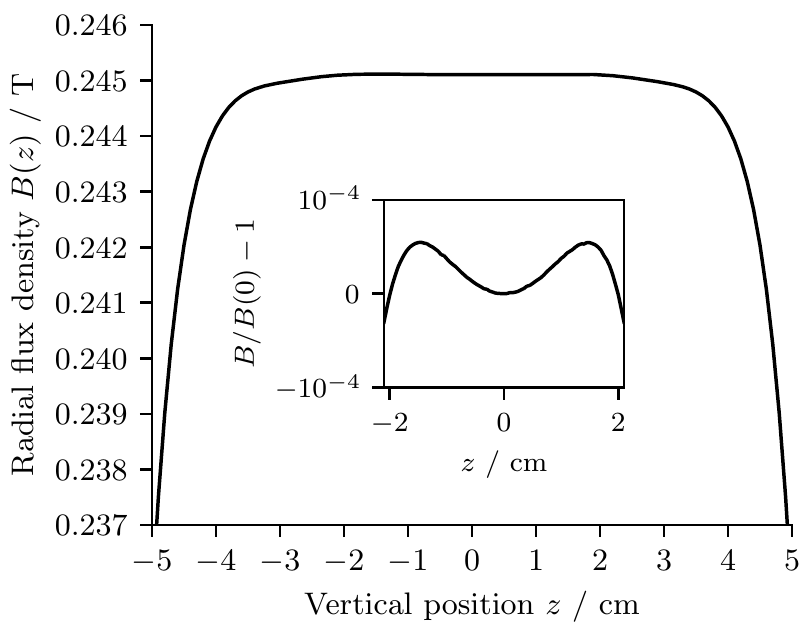}
    \caption{Profile for the radial magnetic flux density $B$. The box inside the plot shows the profile deviation for the precision air gap of \SI{4}{\centi\metre}. This profile was obtained by using finite element analysis.}
    \label{fig:Bl_profile}
\end{figure}

The functional dependence of the radial magnetic flux density on the vertical position is commonly known as profile of the magnetic field. It can be determined by using finite element analysis (FEA). Figure~\ref{fig:Bl_profile} shows a FEA result. The magnetic flux density stays relatively within a band of  \SI{\pm 6e-5}{} in a region of \SI{\pm 4}{cm} about the symmetry plane. For the QEMMS Kibble balance, a stability better than \SI{1e-4}{} is required.
It is possible to decrease the fluctuation of the profile within the \SI{\pm 4}{cm}, by increasing the height $h_\mathrm{m}$ as was observed with FEA. The disadvantage would be a taller and heavier magnet. Since the flatness of the profile is already better than is required, this option was not pursued. Increasing the height $h_\mathrm{m}$ above the optimal height, the flatness of the field will decrease.

The profile flatness has been considered in previous publications \cite{Schlamminger_2012,Seifert_2014,Li_2017_2,You_2017}. Because of the importance of a stable $B$ profile, there are several methods available for optimizing and shimming  the magnet system. Due to the complexity of the problem, it is not simple to obtain an analytical solution. For this reason, most of the publications are based on simulation and measurement results.

\subsection{Coil parameters}

It is necessary to determine the number of the turns and the diameter of the current-carrying coil used with the permanent magnet system. For a given radial magnetic flux density, the length of the wire will determine $Bl$. As described in \cite{Schlamminger_2012}, this product can be chosen in a way to minimize the uncertainty for the mass measurement. This is done by considering the single uncertainties for the measurements of electrical resistance, voltage, velocity and acceleration of free fall, together with the uncertainty equation for the mass measurement:
\begin{equation}
    \frac{\sigma_m^2}{m^2} = \frac{\sigma_{U_R}^2B^2l^2}{R^2m^2g^2} + \frac{\sigma^2_R}{R^2} + \frac{\sigma_g^2}{g^2} + \frac{\sigma_v^2}{v^2} + \frac{\sigma_{U}^2}{B^2l^2v^2}
\end{equation}

\begin{figure}
    \centering
    \includegraphics{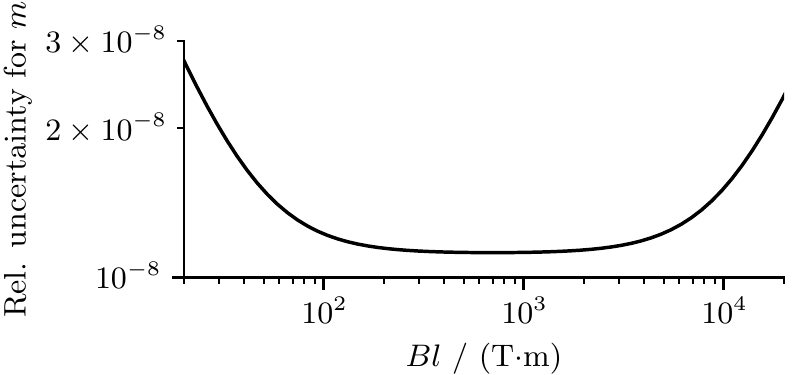}
    \caption{Determination of the $Bl$ value using the same procedure described in \cite{Schlamminger_2012}. There is a flat region with minimum values between \SI{400}{Tm} and \SI{1000}{Tm}.}
    \label{fig:Bl_calc}
\end{figure}

By using the expected uncertainties with this equation it is possible to obtain the relative uncertainty for the mass measurement as a function of the parameter $Bl$. A plot with this relation is shown in figure~\ref{fig:Bl_calc}. The velocity $v$ is expected to be measured with a relative uncertainty of \SI{8e-9}{} for a constant value of about \SI{2}{\milli\metre\per\second}. A resistance $R$ of \SI{1}{\kilo\ohm} will be used for the current measurement and the resistance value is expected to be known with a relative uncertainty of \SI{6e-9}{}. The acceleration of free fall $g$ is measured with a relative uncertainty of \SI{5e-9}{} and the voltages $U_\mathrm{R}$ and $U$ are measured with a uncertainty of \SI{1}{\nano\volt}. The relative uncertainty for the mass measurement was determined for a mass $m$ of \SI{100}{\gram}. A flat region with minimal uncertainties is present for a $Bl$ between \SI{400}{Tm} and \SI{1000}{Tm}. A value of \SI{700}{Tm} is chosen for the QEMMS magnet system. For a radial flux density of \SI{0.24}{T} and a coil radius of \SI{10}{cm}, about \SI{4642}{} turns are necessary to reach this $Bl$.

\begin{figure}
    \centering
    \includegraphics{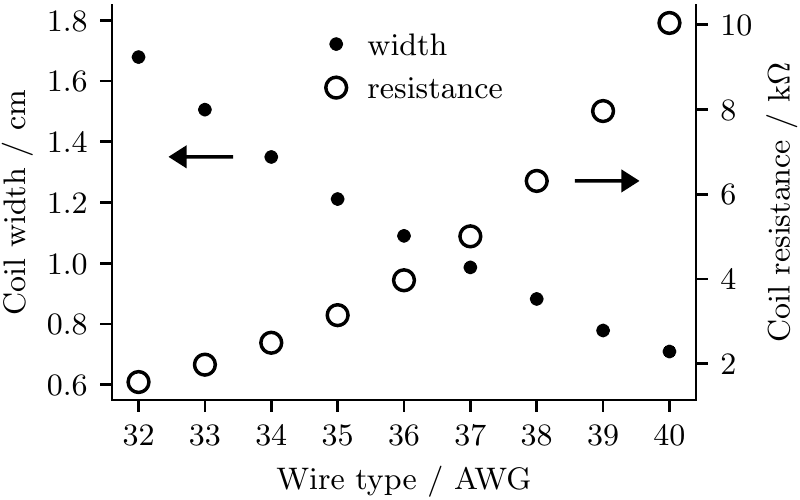}
    \caption{Coil side length and resistance as a function of the wire gauge. A square coil cross section was used for the calculations.}
    \label{fig:coil_width_resistance}
\end{figure}

Once the length of the wire, i.e., the number of turns, is determined, the wire diameter must be chosen. Two competing factors must be weighed against each other. Using a wire with a larger diameter yields a lower resistance for the coil and, hence, less power dissipation and therefore a smaller temperature increase when switching from velocity mode to weighing mode. On the other hand, winding the coil with a wire that has a larger diameter increases the size and weight of the coil. The width of the air gap limits the size of the coil in one direction and plenty of clearance between the former and the iron of the yoke should be taken into consideration to avoid collisions of the coil with the yoke by parasitic motions. The coil width and resistance as a function of the wire type are shown in figure~\ref{fig:coil_width_resistance}. For the results shown in the figure, a square cross section for the coil was assumed, round magnet wires with double insulation layer were considered \cite{MWS_2016} and a packing factor of 0.785 for the winding was assumed. It should be noted that a packing factor of up to 0.86 can be achieved~\cite{McLyman_2004}  for the wire types shown in figure~\ref{fig:coil_width_resistance}. Hence, the calculation is conservative.

\begin{figure}
    \centering
    \includegraphics{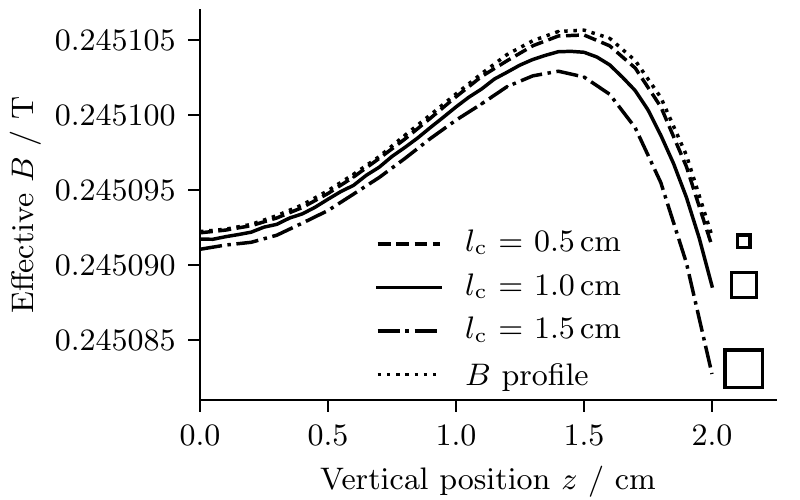}
    \caption{Simulation results for determination of the $B$ profile for different coil side lengths. A square coil cross section was used for the simulations. The coil cross section is shown on the right hand side.}
    \label{fig:coil_side_paper}
\end{figure}

Since the coil has a finite size, the effective profile seen by the coil is an integration of the magnetic flux density over the coil volume. Figure~\ref{fig:coil_side_paper} compares the effective profile of three square coils with different sizes to the original profile.
The curves shown in the figure were obtained with finite element analysis and the mean radius of the  radius $r_\mathrm{c}$ was the same for all calculations.
For all three cases the effective profile is attenuated from the original profile. The larger the square coil, the smaller the effective field. The relative difference between the effective profiles and the original profile, however, are smaller than \SI{4e-5}{} in all three cases. 

\begin{figure}
    \centering
    \includegraphics{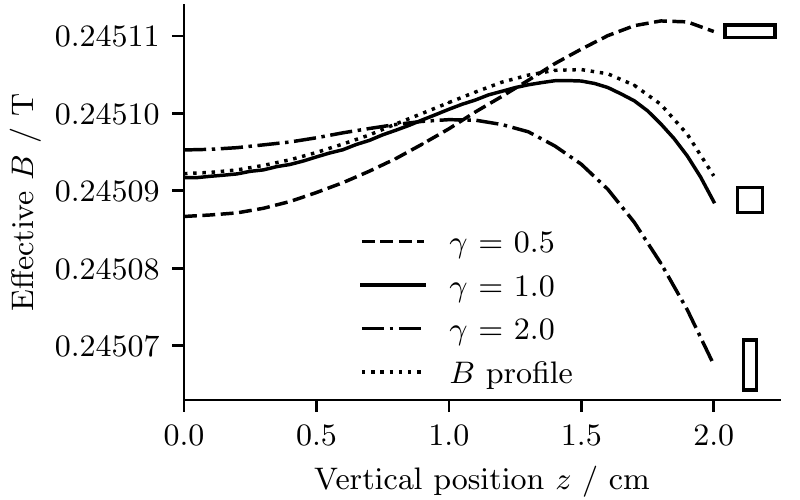}
    \caption{Influence of the aspect ratio in the $B$ profile. For these simulation results a coil with cross section area of \SI{1}{\centi\metre^2} was used. The coil cross section is shown on the right hand side.}
    \label{fig:plot_coil_aspect_ratio}
\end{figure}

Figure~\ref{fig:plot_coil_aspect_ratio} shows the influence of the coil aspect ratio in the flux density profile as observed by the coil. For these simulation results, a coil with cross section area of \SI{1}{\centi\metre^2} was used. The height and width are given as $\sqrt{A_\mathrm{c}}\cdot\gamma$ and $\sqrt{A_\mathrm{c}}/\gamma$ respectively, where $A_\mathrm{c}$ represents the cross section area and $\gamma$ the aspect ratio. As shown in figure~\ref{fig:plot_coil_aspect_ratio}, a higher aspect ratio represents a flatter flux density profile near to the center of the magnet system. However, in this case the deviations are bigger for positions far from the center. For aspect ratios smaller than one, the flatness of the flux density profile in the center is compromised, and the overall deviations in the profile are higher. Aspect ratios equal to one or slightly higher seem the best choices for the Kibble balance application. In NIST-4, for example,  an aspect ratio of 1.12 was chosen. For the QEMMS Kibble balance an aspect ratio of one will be used. The coil will be wound using AWG 36 wires with double insulation leading to a coil resistance of \SI{4}{\kilo\ohm}. In force mode the coil will dissipate \SI{2.1}{\milli\watt}, which is less than half of the power dissipation in the NIST-4.  A coil side length of about \SI{1.1}{\centi\metre}, which is shown in figure~\ref{fig:coil_width_resistance}, represents enough space in the air gap to mount a coil form and operate the balance.

\section{Proposed design of the QEMMS magnet}

The values of the key parameters of the QEMMS magnet system and coil are shown in table~\ref{tab:features_QEMMS}. For comparison the corresponding values of \mbox{NIST-4} are also shown. The major difference between both balances is the measuring range:  QEMMS is being designed to measure masses up to \SI{100}{\gram} with relative uncertainties lower than \SI{2e-8}{}, while \mbox{NIST-4} can measure masses up to \SI{2}{\kilo\gram}, but nominally operates at \SI{1}{\kilo\gram}. QEMMS is intended to be smaller than \mbox{NIST-4} and easier to operate. Hence, a smaller magnet system that is about eight times lighter than that of \mbox{NIST-4} will be used. The split forces are much smaller and an integrated magnet splitter will be used to separate the magnet system in situ and access the coil. TC-16 will be used as the active magnetic material resulting in better temperature stability for the remanence  $B_\mathrm{r}$ than the samarium-cobalt used in \mbox{NIST-4}. The trade-off for a lower temperature coefficient is a lower remanence  which reduces the flux density in the air gap. A smaller travel range equal to \SI{4}{\centi\metre} is required for the QEMMS, and a smaller coil with half of the size of the coil employed in \mbox{NIST-4} will be used. The coil in QEMMS has five times more turns than the coil in \mbox{NIST-4} and, hence, a smaller wire gauge is necessary. As a consequence, the resistance and inductance of the QEMMS coil are larger than the corresponding values in the \mbox{NIST-4} coil.  However, due to the smaller nominal mass in QEMMS (\SI{100}{\gram} vs.  \SI{1}{\kilo\gram}), the thermal power generated in the coil is smaller for QEMMS. Also the reluctance force does not yield a significant contribution to the measurement uncertainty. Although the reluctance force constant is 19 times larger for the QEMMS, the smaller nominal mass means a factor of 1.9 in relative increase. Additionally, the lower mass value causes a smaller deviation in the coil position between the mass-on and mass-off measurement phases described in section~\ref{sec:rel_force}. This behavior also reduces the measurement deviations caused by the reluctance force.

\begin{table}
    \centering
\caption{Comparison between the magnet systems and coils for the QEMMS and the NIST-4.}
    \resizebox{236.6pt}{!}{
    \begin{tabular}{lrr}
    \toprule
     & QEMMS & NIST-4 \\ 
    \midrule
    \multicolumn{3}{l}{\bf Design goals for the balance}\\
    Nominal mass value & \SI{100}{\gram} & \SI{1}{\kilo\gram} \\
    Relative uncertainty & \SI{2e-8}{} & \SI{1e-8}{} \\
    \midrule
    \multicolumn{3}{l}{\bf Parameters of the magnet system}\\
    Flux density $B$ & \SI{0.24}{T} & \SI{0.55}{T} \\
    Precision air gap & \SI{4}{cm} & \SI{8}{cm} \\
    Magnetic material & TC-16 & $\mathrm{Sm_2Co_{17}}$\\    
    Mass of magnet& \SI{110}{kg} & \SI{850}{kg}  \\
    Split force & \SI{250}{N} & \SI{4.7}{kN} \\
    \midrule
    \multicolumn{3}{l}{\bf Parameters of the coil}\\
    Mean radius & \SI{10}{cm} & \SI{21.7}{cm} \\
    Aspect ratio & \SI{1}{} & \SI{1.12}{} \\
    Cross sectional area & \SI{1}{\centi\metre^2} & \SI{2.64}{\centi\metre^2}  \\
    Number of Turns & 4642 & 945 \\
    Wire size & AWG 36 & AWG 24 \\
    Resistance & \SI{4}{\kilo\ohm} & \SI{108}{\ohm} \\
    \midrule
    \multicolumn{3}{l}{\bf Properties of the coil in the magnet }\\
    Inductance & \SI{33.9}{H} & \SI{4.06}{H} \\
    Heating power & \SI{2.1}{\milli\watt} & \SI{5.5}{\milli\watt} \\
    Reluct. force const. & \SI{-4.51}{\per\metre\per\ampere} & \SI{-0.237}{\per\metre\per\ampere} \\
    \bottomrule
    \end{tabular}}
    \label{tab:features_QEMMS}
\end{table}

\section{Summary}

The design considerations and the final design of the magnet system driven by the basic requirements of the Quantum Electro-Mechanical Metrology Suite (QEMMS) were described in this paper. The new magnet system is based on the NIST-4 magnet and it was designed to take advantage of the past performance of NIST-4 while overcoming known practical limitations such as the large temperature coefficient. Analytical models for describing the magnetic flux density in the air gap, the reluctance force in the coil, and the split forces for the separation operation are given in this article. With these models and finite element analysis the performance of the magnet system was evaluated. Aspects related to the coil geometry and flatness of the flux density profile were also considered. The proposed design will be manufactured and tested to verify that the magnet meet its operational requirements.

\appendices

\section{Equations for the magnetic circuit} \label{appx:magnetic_circuit}

\begin{figure}
    \centering
    \includegraphics{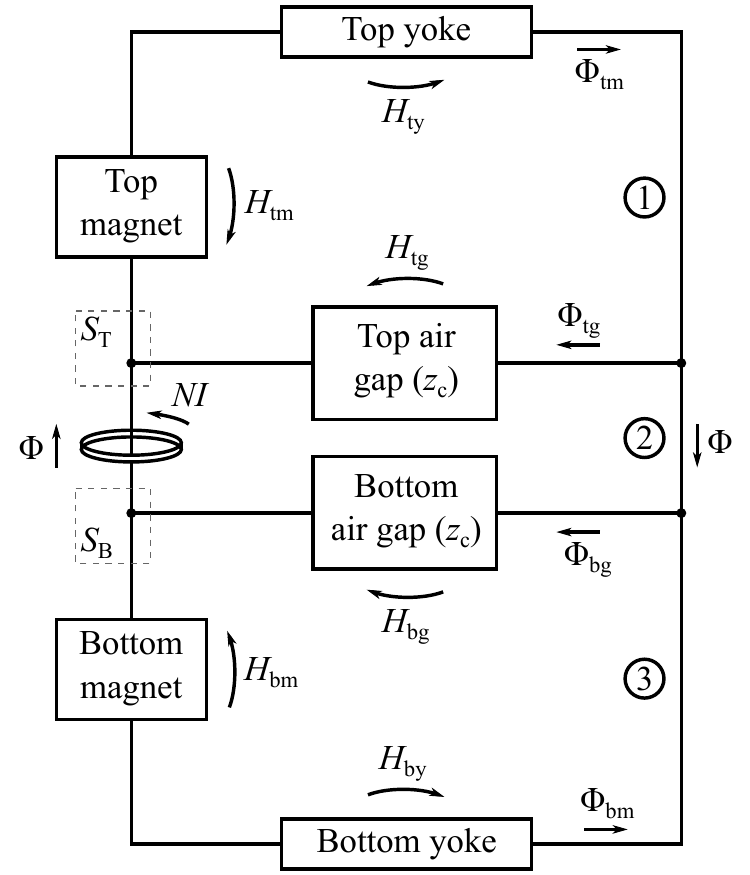}
    \caption{Drawing of the magnetic circuit used to obtain the equations for the system. Top air gap and bottom air gap depend on the coil position $z_\mathrm{c}$.}
    \label{fig:drawing-magnetic_circuit}
\end{figure}

The equivalent magnet circuit for the magnet system of figure~\ref{fig:dimensions_magnet_system} is shown in figure~\ref{fig:drawing-magnetic_circuit}. There are two permanent magnets in the system: the top and bottom magnets. The yoke and the air gap are also divided in two parts named top and bottom. The magnetic flux through the top yoke and top magnet is named $\Phi_\mathrm{tm}$, and the magnetic flux through the air gap is named $\Phi_\mathrm{tg}$. For the lower part, the magnet fluxes $\Phi_\mathrm{bm}$ and $\Phi_\mathrm{bg}$ are defined. The magnetic flux through the coil is given by $\Phi$. The following relationship can be obtained for the magnetic fluxes by applying the Gauss law of magnetism to two separate closed surfaces $S_\mathrm{T}$ and $S_\mathrm{B}$ comprising the nodes above and below the coil respectively:
\begin{equation}
    \Phi = \Phi_\mathrm{tm} - \Phi_\mathrm{tg} = \Phi_\mathrm{bg} - \Phi_\mathrm{bm}
\end{equation}

By using the Ampere circuital law, the following equations can be obtained for the parts of the magnet circuit indicated with the numbers 1, 2 and 3 respectively:
\begin{equation}
    H_\mathrm{ty}\delta_\mathrm{y} + H_\mathrm{tg}\delta_\mathrm{g} - H_\mathrm{tm}\delta_\mathrm{m}=0
\end{equation}
\begin{equation}
    H_\mathrm{bg}\delta_\mathrm{g} - H_\mathrm{tg}\delta_\mathrm{g} = NI
\end{equation}
\begin{equation}
    H_\mathrm{by}\delta_\mathrm{y} + H_\mathrm{bg}\delta_\mathrm{g} - H_\mathrm{bm}\delta_\mathrm{m}=0
\end{equation}
where $H_\mathrm{ty}$ and $H_\mathrm{by}$ are the magnitude of the magnetic field  in the top and bottom yokes, $H_\mathrm{tg}$ and $H_\mathrm{bg}$ are the magnitude of the magnetic field in both parts of the air gap and $H_\mathrm{tm}$ and $H_\mathrm{bm}$ are the magnitude of the magnetic field in the permanent magnets. The quantities $\delta_\mathrm{g}$ and $\delta_\mathrm{m}$ are the width of the air gap and height of the permanent magnets respectively. The effective length of the yoke is defined by $\delta_\mathrm{y}$. It is known that the magnetic field is a vectorial field and, inside the different components of the magnet system, there is a significant variation of this quantity over the space. To obtain the equations above, it is necessary to assume that the magnetic field has a constant magnitude for the yoke, the permanent magnets and the air gap. The quantities that represent the magnitude of the magnetic field can be seen as a mean value along the integration path. 

Since the yoke is made of a ferromagnetic material, the following expressions are obtained for the magnetic flux densities as a function of the magnetic field:
\begin{equation}
    B_\mathrm{ty}=\mu_\mathrm{y}H_\mathrm{ty},\quad B_\mathrm{by}=\mu_\mathrm{y}H_\mathrm{by}
\end{equation}
where the quantity $\mu_\mathrm{y}$ represents the yoke permeability. A linear magnetization curve was assumed for the yoke material.

The permanent magnets are made of rare earth materials and the magnetization curves are also assumed to have a linear behavior:
\begin{equation}
    B_\mathrm{tm}=B_\mathrm{r} - \mu_0\mu_\mathrm{m}H_\mathrm{tm}, \quad B_\mathrm{bm}=B_\mathrm{r} - \mu_0\mu_\mathrm{m}H_\mathrm{bm}
\end{equation}
where $B_\mathrm{r}$ represents the remanence  and  $\mu_\mathrm{m}$  the recoil permeability.

The flux density in both parts of the air gap is given by the following expressions:
\begin{equation}
    B_\mathrm{tg} = \mu_0 H_\mathrm{tg}, \quad B_\mathrm{bg} = \mu_0 H_\mathrm{bg}
\end{equation}

By assuming that the magnet flux density has a constant magnitude inside the yoke, air gap and permanent magnets, the following expressions can be obtained for the magnetic fluxes:
\begin{equation}
    \Phi_\mathrm{tm} = B_\mathrm{ty}A_\mathrm{y}=B_\mathrm{tm}A_\mathrm{m}
\end{equation}
\begin{equation}
    \Phi_\mathrm{bm} = B_\mathrm{by}A_\mathrm{y}=B_\mathrm{bm}A_\mathrm{m}
\end{equation}
\begin{equation}
    \Phi_\mathrm{tg} = B_\mathrm{tg}A_\mathrm{tg}
\end{equation}
\begin{equation}
    \Phi_\mathrm{bg} = B_\mathrm{bg}A_\mathrm{bg}
\end{equation}
where $A_\mathrm{y}$ is the effective area of the yoke. The area of the permanent magnets is given by
\begin{equation}
    A_\mathrm{m}=\pi(r_\mathrm{m}^2-r_\mathrm{i}^2)
\end{equation}
and the areas of the top and bottom parts of the air gap are given by:
\begin{equation}
    A_\mathrm{tg} = 2\pi r_\mathrm{c}(h_\mathrm{m}-z_\mathrm{c})
\end{equation}
\begin{equation}
    A_\mathrm{bg} = 2\pi r_\mathrm{c}(h_\mathrm{m}+z_\mathrm{c})
\end{equation}

By combining the expressions above and neglecting the magnetic field in the yoke, the following expression can be obtained for the magnetic flux through the coil:
\begin{align}\nonumber
\Phi & = \frac{NI\mu_0}{2\delta_\mathrm{g}}( 2 \pi r_\mathrm{c} h_\mathrm{m} + \mu_\mathrm{m}\pi(r_\mathrm{m}^2-r_\mathrm{i}^2)\delta_\mathrm{g} /\delta_\mathrm{m})   \\  \label{eq:mag_flux}
    & + \frac{2\pi r_\mathrm{c}z_\mathrm{c} B_\mathrm{r}}{2 r_\mathrm{c}h_\mathrm{m}/(r_\mathrm{m}^2-r_\mathrm{i}^2) + \mu_\mathrm{m}\delta_\mathrm{g}/\delta_\mathrm{m}}  \\ \nonumber
    & -\frac{\pi r_\mathrm{c} N I z_\mathrm{c}^2 \mu_0 / \delta_\mathrm{g}}{h_\mathrm{m}+\mu_\mathrm{m}(r_\mathrm{m}^2-r_\mathrm{i}^2) \delta_\mathrm{g}/(2 r_\mathrm{c} \delta_\mathrm{m})}
\end{align}

The magnetic field in the yoke can be neglected due to the relative high permeability of the yoke material. This equation has basically three components that are related to the coil inductance in the center of the magnet system, the $Bl$ factor and the reluctance force respectively.

The approach used in this appendix to determine the flux density through the coil is similar to methods used in \cite{Schlamminger_2012,Li_2013,Li_2018,Leupold_1996}.

\section{Equation for the split force} \label{appx:split_force}

The determination of the split force is necessary to design the magnet system in a way to integrate a simple magnet splitter. This force can be determined by integrating the Maxwell stress tensor $\boldsymbol{T}$ along the surface $S$ indicated in figure~\ref{fig:drawing-split-plane}:
\begin{figure}
    \centering
    \includegraphics{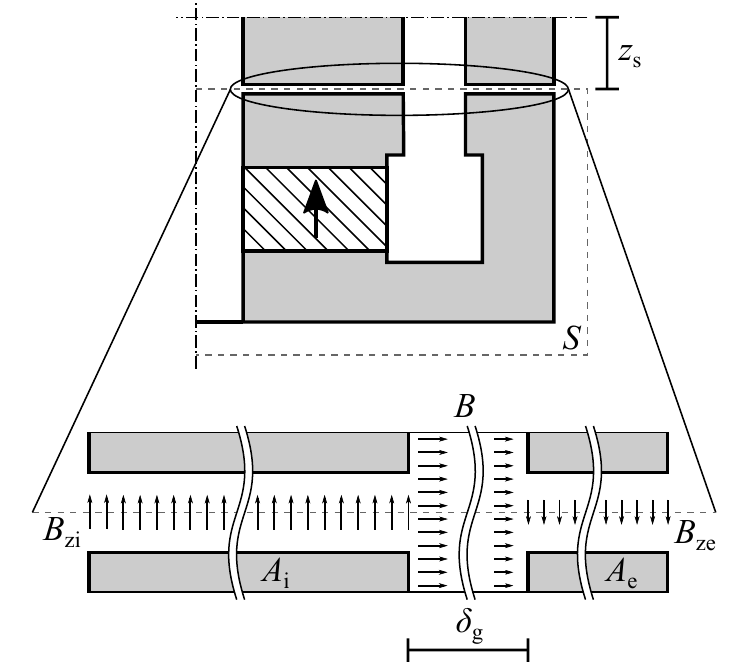}
    \caption{Magnet system and surface area $S$ used for integration of the Maxwell stress tensor and determination of the split force.}
    \label{fig:drawing-split-plane}
\end{figure}
\begin{equation}
    \boldsymbol{F} = \oiint_{S} \boldsymbol{T} \cdot \mathrm{d}\boldsymbol{a}
\end{equation}
where the Maxwell stress tensor is given by:
\begin{equation}
    T_{ij}=\frac{1}{\mu_0}\left( B_i B_j - \frac{1}{2} \delta_{ij}B_\mathrm{m}^2 \right)
\end{equation}

These equations were obtained by assuming static behavior and the absence of electric fields. The separation between both parts of the magnet system is assumed to be very small. A derivation for this equation is described with details in \cite{Griffiths_1990}. In the original derivation a Cartesian coordinate system was used and $B_i$ or $B_j$ represent the magnet flux densities along the different directions. That means $i$ and $j$ can be equal to the directions $x$, $y$ and $z$. The quantity $B_\mathrm{m}$ represents the magnitude of the flux density and $\delta_{ij}$ is the Kronecker delta. For cylindrical coordinates the following equation can be obtained:
\begin{equation} \label{eq:F_z}
    F_{\mathrm{z}} = \frac{1}{2\mu_0}(B_{\mathrm{ze}}^2 A_{\mathrm{e}} + B_{\mathrm{zi}}^2 A_{\mathrm{i}} - 2 \pi r_{\mathrm{c}} B^2  \delta_{\mathrm{g}})
\end{equation}
where $B_\mathrm{ze}$ and $B_\mathrm{zi}$ are the magnetic flux densities shown in figure~\ref{fig:drawing-split-plane}. Due to the symmetry of the problem, for the split plane in the middle, that means $z_\mathrm{s}=0$, the flux densities $B_\mathrm{ze}$ and $B_\mathrm{zi}$ are equal to 0. The flux density is horizontal in the air gap and vertical in the split plane. The flux densities $B_\mathrm{ze}$ and $B_\mathrm{zi}$ are proportional to the height of the split plane $z_\mathrm{s}$ and $B$.
\begin{equation} \label{eq:B_ze}
    B_\mathrm{ze} = 2\pi r_{\mathrm{c}} z_\mathrm{s} B/A_{\mathrm{e}}
\end{equation}
\begin{equation} \label{eq:B_zi}
    B_\mathrm{zi} = 2 \pi r_{\mathrm{c}} z_\mathrm{s} B/A_{\mathrm{i}}
\end{equation}
with the areas $A_\mathrm{i}$ and $A_\mathrm{e}$ given as:
\begin{eqnarray} \label{eq:area_ai}
    A_\mathrm{i} = \pi (r_{\mathrm{c}}-\delta_\mathrm{g}/2)^2 - \pi r_\mathrm{i}^2 \\ \label{eq:area_ae}
    A_\mathrm{e} =  \pi r_\mathrm{e}^2 - \pi (r_{\mathrm{c}} + \delta_\mathrm{g}/2)^2
\end{eqnarray}

In order to obtain these equations it was assumed that the magnetic flux density outside the separation region is near to zero. A combination of (\ref{eq:F_z}), (\ref{eq:B_ze}) and (\ref{eq:B_zi}) gives the following expression for determination of the split force:
\begin{equation}
    F_\mathrm{z}=\frac{\pi r_{\mathrm{c}} B^2}{\mu_0} \left(  \frac{2\pi r_\mathrm{c} (A_\mathrm{i} + A_\mathrm{e})}{A_\mathrm{i} A_\mathrm{e}}z_\mathrm{s}^2 - \delta_{\mathrm{g}} \right)
\end{equation}

\bibliographystyle{IEEEtran}

\bibliography{bibliography}

\end{document}